\documentclass[12pt]{article}
\usepackage{latexsym}
\usepackage{graphicx}

\thicklines                         
\long \def \blockcomment #1\endcomment{}

\begin{document}           
\baselineskip=0.33333in

\vglue 0.5in
\begin{center}{\bf Remarks on the Proton Structure}
\end{center}
\begin{center}E. Comay$^*$
\end{center}

\begin{center}
Charactell Ltd. \\
PO Box 39019 \\
Tel-Aviv, 61390 Israel
\end{center}
\vglue 0.5in
\noindent
\vglue 0.2in
\noindent
Abstract:

Elastic and inelastic cross section of proton-proton and electron-proton
scattering are discussed. Special attention is given to
elastic scattering and to the striking
difference between the data of these two kinds of experiments. It is
shown that the regular charge-monopole theory explains the main
features of the data. Predictions of results of CERN's
Large Hadron Collider are pointed out.

\newpage
\noindent
{\bf 1. Introduction}
\vglue 0.33333in
Scattering experiments are used as a primary tool for investigating
the structure of physical objects. These experiments can be divided
into several classes, depending on the kind of the colliding particles.
The energy involved in
scattering experiments has increased dramatically
during the century elapsed since the celebrated Rutherford experiment
has been carried out. Now, the meaningful
value of scattering energy is the quantity measured in the
rest frame of the projectile-target center of mass.
Therefore, devices that use colliding beams enable measurements of
very high energy processes. For this reason, the Large Hadron Collider
(LHC) facility
at CERN, which is designed to produce 14 TeV proton-proton ($pp$)
collisions, will make a great leap forwards.

This work examines two different scattering data of protons. One set
consists of the pre-LHC $pp$ scattering data and the second set is
electron-proton and positron-proton scattering data. Hereafter, $ep$ denote
these lepton-proton scattering experiments.
A special attention is given to elastic scattering (ES), where the proton
remains intact and no new particle
is produced. The data prove that the elastic cross section (ECS)
of $pp$ scattering differs dramatically from that of $ep$.
These experimental data are explained by the Regular Charge-Monopole
Theory (RCMT). It is also shown how this theory together with currently
available data yields a prediction of LHC results.

The Lorentz metric used is diagonal and its entries are (1,-1,-1,-1).
Expressions are written in units where $\hbar = c = 1$. In this
system of units there is just one dimension. Here it is taken to be
that of length. Therefore, the dimension of a physical quantity is a
power of length and is denoted by $[L^n]$.

The data on cross sections of $pp$ and $ep$ are presented
in the second Section. An analysis of these data and a theoretical
explanation of their main features are included in the third Section.
The fourth Section discusses the structure of the baryonic core.
The last Section contains concluding remarks.

\vglue 0.66666in
\noindent
{\bf 2. The Relevant Cross Section Data}
\vglue 0.33333in

   Let us turn to the $pp$ scattering
(see fig. 1 and its original version [1]). Points $A,\;B,\;C$
divide the ECS graph into four parts. For a laboratory
momentum smaller than that of point
$A$, the elastic cross section shows the
characteristic decreasing pattern of a Mott-like scattering.

\begin{figure}[h]
  \centering
    \includegraphics[width=10cm]{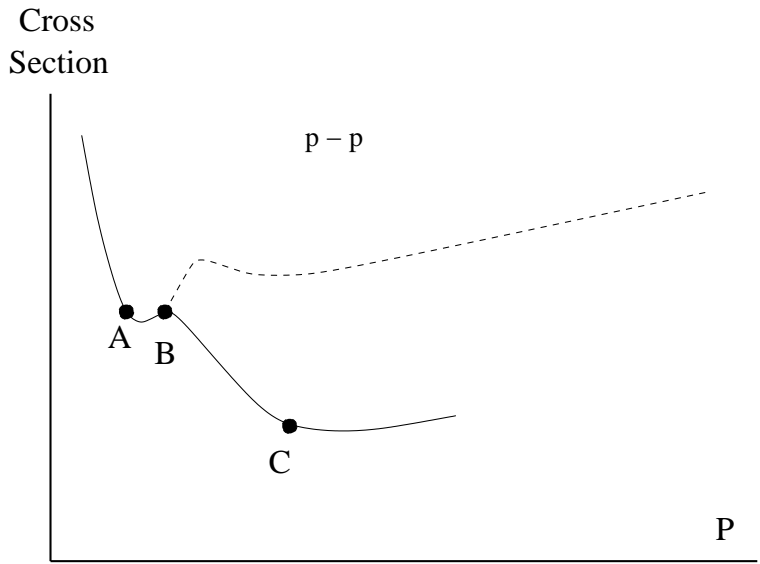}
  \caption{\em Proton-proton cross section versus the laboratory momentum P.
Axes are drawn in a logarithmic scale.
The continuous line denotes
elastic cross section and the broken line denotes total cross section.
Points $A,B,C$ help the discussion (see text).
(The accurate figure can be found in [1]).}
\end{figure}

Clearly, the Mott-like decrease of the cross section does not
hold for a momentum greater than that of point $A$. For the momentum
interval $[A,B]$, a new force enters the scattering process.
This is the nuclear force whose phenomenological properties are well
known for a very long time [2]. Its main features are a quite strong
repulsive force at the nucleon's
inner part and an attractive force outside it.
The nuclear attractive force decays more rapidly then the Coulomb force. At
a short distance from the proton's center,
these forces are much stronger then the electromagnetic force.
(The fact that near the origin
the potential of the nuclear force $V(r)$ varies
more rapidly than the potential of the Coulomb force $1/r$, is
significant. This point is discussed in the third Section.)
For momentum values belonging to the interval $[A,B]$, the
nuclear force alters the direction of the graph that describes
ECS. Here the decrease of ECS stops continuously and for a certain
interval of the projectile's momentum, ECS increases with it.

Let us examine the momentum interval belonging to points $[B,C]$. Fig. 1
indicates that a new process begins to take place
for momentum values greater than that of point $B$.
For these values, the collision's energy is large enough
for producing hadrons. It means that inelastic scattering
begins at point $B$. The inelastic cross section (ICS)
is the difference between the broken line describing
the total cross section (TCS) and the continuous line describing ECS.
Thus, for momentum values greater than that of point $B$, ECS begins
to decrease. An examination of the scale of the
original figure [1] indicates that ICS becomes
greater then ECS and at point $C$, ICS is about five times greater then ECS.

   For momentum values greater than that of point $C$, the
decreasing pattern of ECS gradually stops
and it slightly begins to increase together with the momentum.
It can be concluded that points $A,B,C$ of the graph
show clearly four momentum regions,
each of which has a unique behavior of ECS.

In the second kind of scattering data, one proton is replaced by an
electron. Unfortunately, in the case of $ep$ scattering, publications
of Particle Data Group, like
[1], do not contain a figure which is analogous
to fig. 1. Therefore, the discussion relies on appropriate formulas
that describe the data. The following arguments prove that in $ep$
scattering, the characteristics of the cross section differ
substantially from the $pp$ data depicted in fig. 1.

Let us examine the elastic $ep$ scattering.
In this case the analysis uses the Rosenbluth formula. Here
the Mott cross section is factored out and is multiplied by trigonometric
functions and form factors which depend on the square of the
4-momentum transferred $q^2$.
The Mott differential cross section takes the following form
(see [3], p. 192)
\begin{equation}
(\frac {d\sigma }{d\Omega })_{Mott} =
\frac {\alpha ^2 \cos ^2 (\theta /2)}
{4p^2 \sin ^4 (\theta /2)[1 + (2p/M) \sin^2 (\theta /2)]},
\label{eq:MOTT}
\end{equation}
where $\alpha \simeq 1/137$ denotes the square of the electron's charge
and $p$ is the linear momentum of the incoming electron.
The Rosenbluth formula can be cast into the following form
(see [3], p. 193, eq. (6.26))
\begin{equation}
(\frac {d\sigma }{d\Omega })_{Rosenbluth} =
(\frac {d\sigma }{d\Omega })_{Mott}
(A(q^2) + B(q^2) \tan ^2(\theta /2)),
\label{eq:ROSENBLUTH}
\end{equation}
where $A$ and $B$ are related to the proton's electric and the magnetic
form factors.

In the $ep$ case,
the Rosenbluth formula $(\!\!~\ref{eq:ROSENBLUTH})$ is used for
describing ECS. This formula is valid
even for the very high energy of modern
colliders [4]. A useful quantity is called the dipole
form factor (see [3], p. 196)
\begin{equation}
G_D(q^2) = (1 + q^2/0.71)^{-2},
\label{eq:DIPOLE}
\end{equation}
where $q^2$ is measured in GeV$^2$. It turns out that,
for all energies used up to date, the actual form
factors are nearly the same as $(\!\!~\ref{eq:DIPOLE})$. For the
present discussion it is sufficient to realize that
the actual form factors are of the same order of magnitude as
$(\!\!~\ref{eq:DIPOLE})$ (see figures 2,3 of [5]). It can be
concluded that the $ep$ elastic form factor decreases with an
increase of $q^2$ and for $q^2 \gg $ 1 GeV$^2$, the contribution of
the electric form factor $G_E$ decreases like $q^{-4}$ and that of
the magnetic form factor decreases like $q^{-2}$ (see [3], p. 193).
Moreover, these decreasing contributions are multiplied by the Mott cross
section that decrease like $p^{-2}$.

The form of the deep inelastic cross section is quite different.
Using the notation of [3], sections 8.2 and 8.3, one finds for deep
inelastic collisions
\begin{equation}
(\frac {d^2\sigma }{dq^2 dx}) =
\frac {4\pi \alpha ^2}{q^4}[(1-y)\frac {F_2(x,q^2)}{x} +
y^2 F_1(x,q^2)].
\label{eq:DEEP}
\end{equation}
Here $x$ and $y$ are dimensionless variables whose value falls
in the range [0,1]. Experimental evidence proves
that Bjorken scaling holds and that the expression inside the square
brackets of $(\!\!~\ref{eq:DEEP})$ is very nearly
independent of $q^2$. Hence, integrating $(\!\!~\ref{eq:DEEP})$
on $q^2$, one finds that the deep inelastic cross section decreases
like $1/q^2$. This is the rate of the decrease of the magnetic
form factor contribution to the elastic scattering.

The first interesting
issue is the ratio of ECS to TCS which is found for very high energy.
As stated above, in the case of $pp$ scattering,
this ratio is about 1/6. On the other hand, the additional $1/p^2$
factor of the Mott cross section $(\!\!~\ref{eq:MOTT})$
proves that in $ep$ scattering,
elastic events are very rare. Hence, the data lead one to
the following conclusion:
\begin{itemize}
\item[{I.}] For very high energy, about $16\%$ of the $pp$ scattering events
are elastic whereas in the corresponding $ep$ scattering, the percentage
of elastic events is very, very small.
\end{itemize}
The second issue is the behavior of ECS as a function of energy in
the two kinds of scattering experiments described above. The data of
fig. 1 (see [1]) shows how the $pp$ cross section varies as a function
of either the projectile's momentum or, equivalently, on a different scale,
as a function of
$\sqrt S$, where $S$ is the square of the invariant energy of the
colliding particles.

The ECS graph of fig. 1 shows that it stops
decreasing for energies which are somewhat greater than that of
point $C$. In the case of the $ep$ scattering, the
ECS information is given in terms of the differential cross section
which depends on the invariant square of the
momentum transferred $q^2$. However, the following calculation
proves that for $ep$ scattering ECS does not stop decreasing with
the increase of the collision's energy.

The calculation is carried out in the rest frame of the colliding
particles. Let the 4-momentum of the incoming electron be
\begin{equation}
p^\mu _{in} = (\sqrt {p^2 + m^2},0,0,p)
\label{eq:P_IN}
\end{equation}
and that of the outgoing electron is
\begin{equation}
p^\mu _{out} = (\sqrt {p^2 + m^2},p\sin \theta,0,p\cos \theta).
\label{eq:P_OUT}
\end{equation}
It follows that the square of the momentum transferred is
\begin{equation}
q^2 = (p^\mu _{in} - p^\mu _{out} )(p_{\mu \,in} - p_{\mu \,out}) =
2p^2(1 - \cos \theta).
\label{eq:Q2}
\end{equation}
Now the elastic cross section $\sigma $ is the spherical integral of the
Rosenbluth differential cross section formula
$(\!\!~\ref{eq:ROSENBLUTH})$
\begin{equation}
\sigma = \int
(\frac {d\sigma }{d\Omega })_{Rosenbluth} \sin \theta d\theta d \phi.
\label{eq:SIGMA}
\end{equation}

Let us examine the integrand of $(\!\!~\ref{eq:SIGMA})$ at a certain
value of $(\theta ,\phi )$. Relation $(\!\!~\ref{eq:Q2})$
proves that a replacement of $q^2$ by $p^2$ is followed
by a trigonometric factor. Now if $p$ increases then the Mott
factor $(\!\!~\ref{eq:MOTT})$ decreases and the same is true for
the dipole factor $(\!\!~\ref{eq:DIPOLE})$. Therefore, since the
differential cross section is positive and it decreases for all
values of $(\theta ,\phi )$, one concludes that the spherical integral
$(\!\!~\ref{eq:SIGMA})$ also decreases with increasing momentum.
This outcome proves that for the $ep$ scattering discussed here,
ECS does not stop decreasing with an increase of the linear momentum.
The validity of the following conclusion relies on these results.
\begin{itemize}
\item[{II.}] For high enough energy
there is a substantial difference between elastic scattering of $pp$
and of $ep$. If the collision energy increases then
ECS of $pp$ stops decreasing and even shows a small
increase whereas ECS of $ep$ does not stop decreasing.
\end{itemize}

There is another kind of difference between $ep$ and $pp$ scattering.
The Mott formula decreases like $1/p^2$. This matter is evident on
the basis of dimensions arguments. The cross section has the dimension
$[L^2]$. In the system of units used here, energy and momentum have the
dimension $[L^{-1}]$ and the coupling constant $\alpha $ is dimensionless.
This argument explains the negative slope of the cross section
$\sigma (p^2)$ of the Mott formula $(\!\!~\ref{eq:MOTT})$
and of the deep inelastic scattering of $ep$ $(\!\!~\ref{eq:DEEP})$.

This property does not hold for the $pp$ scattering. Here one sees
that for low momentum, which is smaller than that of point $A$
of fig. 1, The TCS decreases steeply, as expected from the Mott formula.
Let us turn to momentum
values which are greater then that of point $A$ of fig. 1, For most
of these momentum regions, TCS increases and at a short region of
momentum values, it decreases. However, the decrease rate is much smaller
than the quite steep slope of $1/p^2$. Therefore, it is concluded that
\begin{itemize}
\item[{III.}] Unlike the case of $ep$ scattering, one finds that for
high enough energy, TCS of $pp$ does not follow the $1/p^2$ decrease of
the Mott formula..
\end{itemize}

Physical consequences of conclusions I-III of this Section are
discussed in the rest of this work.

\vglue 0.66666in
\noindent
{\bf 3. A Discussion of the Cross Sections}
\vglue 0.33333in

The $pp$ and $ep$ scattering experiments help us understand the proton's
structure. Conclusions I-III of the previous Section
reveal a dramatic difference between the results
of these experiments. For example, experiments done in the HERA
facility at DESY report on $ep$ collisions where $q^2 > 10000$
GeV$^2$ [6-8]. Substituting these values of $q^2$ in the dipole
formula $(\!\!~\ref{eq:DIPOLE})$, and remembering that the
magnetic form factor dominates very high $q^2$ collisions,
one finds that for $ep$ scattering the
ratio of ECS to ICS is less than $10^{-4}$. On the other hand,
in the $pp$ scattering this ratio is about 1/5. These
properties are important for the discussion carried out below.
Therefore, they are summarized as follows:
\begin{itemize}
\item[{1.}] For nearly all
events of energetic collisions of electrons with
proton's quarks, the proton is broken apart and the fragments come
out as a set of hadrons. The relative number of elastic collisions
where the proton remains intact is very, very small.
\item[{2.}] In $pp$ collision of similar energy, about $16\%$
of the events are elastic and in these cases the proton remains intact.
\end{itemize}

The following explanation of properties 1,2 is adopted here:
\begin{itemize}
\item[{a.}] The proton consists of quarks and of another object called
baryonic core. The existence of the baryonic core is consistent
with the experimental evidence showing
that for an ultra-relativistic proton,
quarks carry just about one half of the proton's linear momentum
(see [3], p. 282).
\item[{b.}] The baryonic core is electrically neutral. Therefore,
electrons do not interact with it. (Conditions for a
possible deviation from this behavior are discussed in the next Section.)
\item[{c.}] The baryonic core participates in strong interactions.
Hence, three kinds of interacting pairs of particles exist in
a very high energy $pp$ collision: quark-quark, quark-core and
core-core.
\item[{d.}] The core is a relatively rigid object and a core-core
interaction is likely to produce an elastic collision.
\end{itemize}

Statements a-d make a phenomenological explanation of the data discussed
in this work in general and of points 1,2 in particular. Thus, in a $pp$
collision there is a core-core interaction. The relative rigidity
of the core is the primary reason for the non-negligible part
of ECS in $pp$ collision. The fact that the core is electrically.
neutral explains why it does not contribute to ECS of $ep$ collisions.
The following lines present a theoretical basis
for points a-d..

It has been proved that one can use very well established physical
principles and construct a regular theory of electric charges and
magnetic monopoles RCMT [9,10]. The main
results of RCMT can be put in the following words:

Charges do not interact with bound fields of monopoles and
monopoles do not interact with bound fields of charges. Charges interact
with all fields of charges and with radiation fields emitted from
monopoles. Monopoles interact with all fields of monopoles and with
radiation fields emitted from charges. Another important result of
RCMT is that the unit of the elementary magnetic charge $g$ is a
free parameter. However, hadronic data indicate that
this unit is much larger than that of the electric charge
$g^2 \gg e^2 \simeq 1/137$.
More details of RCMT can be found in [9-11].

These properties of RCMT
fit like a glove the data of electromagnetic projectiles
interacting with nucleons (see [11], pp. 90-92).
Thus, protons and neutrons do not look alike
in cases of charged lepton scattering whereas they look very similar
if the projectile is a hard enough real photon (see [12] and the figure on
p. 369 of [1]).

Electrodynamics of magnetic monopoles is dual to electrodynamics of
electric charges. This analogy is helpful for understanding the
applicability of RCMT to hadrons in general and to baryons in particular.
Baryons do not show the static force that should be found between
monopoles. Therefore, one
must assume that they are neutral with respect to magnetic charge.
Thus, each quark is assumed to carry one negative unit of monopole.
The overall monopole charge of baryons vanishes because the baryonic
core carries three positive units of magnetic charge. (The relative sign
of the monopole charge of quarks and of the baryonic core
is arbitrary. Here it is defined so that the similarity with the
respective electric charge of electrons and nuclei holds.)
The elementary unit
of magnetic charge is much larger then that of the electric
charge. For this reason,
baryonic quarks are very tightly bound to the
core (provided the comparison is made with atomic electrons).
Therefore, a baryon can be regarded as
a magnetic monopole analog of an atom, where quarks are strongly bound
to the core. A quark is analogous to an electron and the baryonic core is
analogous to the atomic nucleus.

Now let us use this structure of baryons together
with very well established physical principles for an
interpretation of a $pp$ scattering process. The discussion is carried
out in the rest frame of one proton (the target). The projectile
interacts with the static potential of the target. As the linear
momentum of the projectile increases, its wave length decreases
and its wave function changes sign more rapidly. Therefore, spatial
regions where the potential varies slowly make a very
small contribution to the scattering of very high energy.
This general quantum
mechanical argument proves that for a very short wave length
of the projectile, a meaningful contribution to the scattering process
is obtained only from the region near the baryonic core,
where the potential varies strongly.

Another point which is relevant to the discussion is the $1/r$
variation of the Coulomb potential. This kind of variation leads
to the Rutherford and the Mott $(\!\!~\ref{eq:MOTT})$ scattering
formulas, where the cross section falls like $p^{-2}$. Therefore,
a TCS rise of more energetic
$pp$ scattering must be related to a potential whose spatial
variation is stronger than the Coulomb $1/r$ value. This requirement
holds for the two regions of fig. 1 where the cross section increases:
the regions where the momentum is larger than that of points $A,C$,
respectively.

Let us begin with
momentum values which are smaller than that of point $B$ of
fig. 1. For momentum values which are smaller than that of point
$A$ of fig. 1, one finds a typical Mott-like $1/p^2$ decrease of the
cross section. Let us turn to momentum values of the interval
$[A,B]$. Here the nuclear force enters the process.
Relative to the Coulomb force, this force varies
very rapidly at the spatial region where it is not negligible.
Therefore, the increase of TCS between points $[A,B]$ is understood.
For momentum values greater than that of point $C$, a
similar effect is found inside the proton. Indeed, one should
remember that by analogy with atomic structure,
valence quarks screen the core's potential. Therefore, as
a particle (either a quark or the core of the projectile) approaches
the baryonic core of the target, the interaction increases more rapidly
than the $1/r$ rate of a Coulomb potential. Hence, the contribution of
the quark-core and core-core interaction increases. The former
mainly affects ICS and the latter mainly affects ECS.
Thus, for an energy greater than that of point $C$ of fig. 1,
the increase of both TCS and ECS is also understood.

\vglue 0.66666in
\noindent
{\bf 4. The Structure of the Baryonic Core}
\vglue 0.33333in

Up to this point, the discussion relies only the fact that the
baryonic core carries three units of monopole charge. (This property
is mandatory for RCMT, because of the need to
explain the neutrality of
baryons with respect to magnetic charge.) Referring to the problem
of the core's structure, one may consider two alternatives:
\begin{itemize}
\item[{1.}] The core is a simple elementary pointlike object.
\item[{2.}] The core contains closed shells of quarks.
\end{itemize}
The first case is certainly simpler than the second case. However,
one should not expect to find that Nature is too simple. In particular,
there are two different experimental data that support the second case.

The experiments carried out in the HERA facility at DESY [6-8]
report that the number of events of very high $q^2$ $ep$
scattering is more than expected.
This result can be explained as an analog of the
Franck-Hertz experiment: interaction with bound particles takes place only
if the projectile's energy is high enough. In the present discussion, the
bound particles are quarks of closed shells of the baryonic core.
A different kind of data is the charge radius of the proton and that
of the $\pi ^\pm $ meson. It can be shown why these data provide
two different arguments supporting the existence of closed shells of
quarks at the baryonic core [13]. On the basis of these experiments,
it is assumed here that the baryonic core has closed shells
of quarks. Let us see how this assumption is expected to
affect results of CERN's LHC.

The HERA experiments also provide information on the energy required
for exciting quarks belonging to the baryonic core. This statement
relies on [6-8] whose results are explained by this kind of
excitation. At HERA, the proton,
energy is 820 GeV and that of the positron is 27.5 Gev [7]. Obviously,
for these ultra-relativistic values, these numbers also represent the
linear momentum. At LHC, a proton replaces HERA's positron. Here a
rough comparison
of the collision energy of these colliders is described. The
evaluation uses an appropriate replacement of the positron,
which is an elementary pointlike particle, by one of the protons
that participate in the LHC collision. Thus, let
$P$ denote an LHC proton that corresponds to a proton at HERA and
$P'$ denotes an LHC proton that replaces HERA's positron. Now, we know that
in an ultra-relativistic proton, valence quarks together with the
additional $\bar {q}q$ pairs carry about one half of the proton's
momentum (see [3], p. 282). Hence, all quarks of $P'$ carry kinetic
energy of 3500 GeV and each quark of $P'$ carries about 800 GeV of
kinetic energy. (This estimate relies on the fact that beside the three
valence quarks, there is a nonvanishing probability that
the proton contains $\bar {q}q$ pairs.)
These arguments show that at LHC, the energy of the proton $P$ is more
than eight times larger than that of HERA's proton and the energy of each
quark of LHC's $P'$ is about 30 times larger than that of HERA's
positron. It follows that at LHC, collisions with the baryonic
core will be much more energetic than those of HERA.

Relying on this analysis, one can predict that
the number of energetic LHC events will be
more than those which are expected on the basis of the present
knowledge of valence quarks and of the $\bar {q}q$ pair. Obviously,
excited quarks of the core are expected to behave like valence
quarks and contribute mainly to an inelastic process. In other
words, for this gigantic energy, the baryonic core stops
behaving as a rigid object. These
arguments lead to the following predictions of LHC results:

\begin{itemize}
\item[{1.}] For very energetic LHC collisions, more inelastic events will
be found (comparing to the collider's data of [1]).
\item[{2.}] This increase of the number of inelastic events will be
accompanied by a {\em decrease} of the number of elastic events.
\end{itemize}


\vglue 0.66666in
\noindent
{\bf 5. Conclusions}
\vglue 0.33333in

This work discusses the striking difference between high energy $ep$
scattering and $pp$ scattering data. Elastic events are very, very rare
in high energy $ep$ scattering. Therefore it is concluded that if
a valence quark (or a member of a $\bar {q}q$ pair) is hit by an
energetic projectile then an inelastic event follows.

In the corresponding $pp$ scattering, elastic events are about 1/5
of the number of inelastic events. Hence, the relative portion of
elastic events is larger by several orders of magnitude than that of
the corresponding events of $ep$ scattering. Therefore, one must
look for another component included in the proton. This component
must be rigid enough for absorbing the energy exchanged in a high
energy collision without causing a proton disintegration
into a set of hadrons. Since this
component is not detected by electrons (and not by positrons), it must
be electrically neutral. This proton component is called here
the baryonic core.

The existence of a baryonic core is a self-evident
result of the Regular Charge-Monopole Theory. Therefore, the experimental
data discussed in this work provide another support for the
applicability of this theory to hadrons. (See [11] for other
arguments of this kind.)

It is also explained why one should expect that the baryonic core
contains closed shells of quarks. The HERA data indicate that
the LHC energy is higher then the energy required for
exciting quarks of the baryonic core. Therefore, the first prediction
made at the end of the previous Section says
that the very energetic collisions of the LHC will produce more
inelastic events than expected on the basis of present collider
data of valence quarks. A support for this prediction can be
found in the highest part of [1], which is based on cosmic ray experiments.
A second prediction says that the increase of the number of inelastic
events will be followed by a decrease of the number of elastic events.
An LHC confirmation of these predictions will provide another support for
the relevance of the Regular Charge-Monopole Theory to strong
interactions.


\newpage
References:
\begin{itemize}
\item[{*}] Email: elicomay@post.tau.ac.il  \\
\hspace{0.5cm}
           Internet site: http://www.tau.ac.il/$\sim $elicomay
\item[{[1]}] C. Amsler et al., Phys. Lett. {\bf B667}, 1 (2008).
(See p. 364).
\item[{[2]}] A. deShalit and H. Feshbach, {\em Theoretical Nuclear
Physics} (John Wiley, New York, 1974). Vol 1, pp. 11-18.
\item[{[3]}] D. H. Perkins, {\em Introduction to High Energy Physics}
(Addison-Wesley, Menlo Park CA, 1987).
\item[{[4]}] I. A. Qattan et al., Phys. Rev. Lett., {\bf 94}, 142301 (2005).
\item[{[5]}] H. Gao, Int. J. Mod. Phys. E {\bf 12}, 1 (2003).
\item[{[6]}] C. Adloff et al. {\em Z. Phys. C} {\bf 74}, 191 (1997).
\item[{[7]}] J. Breitweg et al. {\em Z. Phys. C} {\bf 74}, 207 (1997).
\item[{[8]}] A. Wagner {\em Tr. J. of Physics} {\bf 22} 525 (1998).
\item[{[9]}] E. Comay {\em Nuovo Cimento B} {\bf 80}, 159 (1984).
\item[{[10]}] E. Comay {\em Nuovo Cimento B} {\bf 110}, 1347 (1995).
\item[{[11]}] E. Comay
{\em A Regular Theory of Magnetic Monopoles and Its Implications} in
{\em Has the Last Word Been Said on Classical Electrodynamics?}
ed. A. Chubykalo, V. Onoochin, A. Espinoza and R. Smirnov-Rueda
(Rinton Press, Paramus, NJ, 2004).
\item[{[12]}] T. H. Bauer, R. D. Spital, D. R. Yennie and
F. M. Pipkin {\em Rev. Mod. Phys.} {\bf 50}, 261 (1978).
\item[{[13]}] E. Comay (unpublished). See:
\nolinebreak {http://www.tau.ac.il/~elicomay/LHC.pdf}

\end{itemize}


\end{document}